\newcommand{\diracslash}[1]{#1\llap{/\kern2pt}}

\newcommand{\be}{\begin{equation}}
\newcommand{\ee}{\end{equation}}
\newcommand{\bea}{\begin{eqnarray}}
\newcommand{\eea}{\end{eqnarray}}
\newcommand{\ba}[1]{\begin{array}{#1}}
\newcommand{\ea}{\end{array}}

\newcommand{\bt}{\begin{tabular}}
\newcommand{\et}{\end{tabular}}

\newcommand{\beas}{\begin{eqnarray*}}
\newcommand{\eeas}{\end{eqnarray*}}

\documentclass[preprint,prd,aps,floats,nofootinbib,floatfix]{revtex4}
\usepackage{graphicx}
\begin{document}

%\date{\today}
\title{Charmonium decay widths in magnetized matter}
%%%%%%%%%%%%%%%%%%%%%%%%%%%%%%%%%%%%%%%%%%%%%%%%%%%%%%%%%%%%%%%%%
\author{Amruta Mishra}
\email{amruta@physics.iitd.ac.in}
\affiliation{Department of Physics, Indian Institute of Technology, Delhi,
Hauz Khas, New Delhi -- 110 016, India}

\author{Amal Jahan CS}
\email{amaljahan@gmail.com}
\affiliation{Department of Physics, Indian Institute of Technology, Delhi,
Hauz Khas, New Delhi -- 110 016, India}

\author{Shivam Kesarwani}
\email{kesarishvam@gmail.com}
\affiliation{Department of Physics, Indian Institute of Technology, Delhi,
Hauz Khas, New Delhi -- 110 016, India}

\author{Haresh Raval}
\email{haresh.ein@gmail.com}
\affiliation{Department of Physics, Indian Institute of Technology, Delhi,
Hauz Khas, New Delhi -- 110 016, India}

\author{Shashank Kumar}
\email{shashank11297@gmail.com}
\affiliation{Department of Physics, Indian Institute of Technology, Delhi,
Hauz Khas, New Delhi -- 110 016, India}

\author{Jitendra Meena}
\email{jm.jitendermeena@gmail.com}
\affiliation{Department of Physics, Indian Institute of Technology, Delhi,
Hauz Khas, New Delhi -- 110 016, India}

\begin{abstract}
We study the partial decay widths of the charmonium states ($J/\psi$, 
$\psi(3686)$, $\psi(3770)$, $\chi_{c0}$, $\chi_{c2}$) 
to $D\bar D$ ($D^+D^-$ or $D^0\bar {D^0}$)
in isospin asymmetric nuclear matter, in the presence of
strong magnetic fields.
The in-medium partial decay widths of charmonium states
to $D\bar D$ are calculated within a light quark--antiquark
pair creation model, namely the $^3P_0$ model,  
using the in--medium masses 
of the charmonia as well as $D$ and $\bar D$ mesons
in the magnetized nuclear matter,
obtained within a chiral effective model.
The presence of a magnetic field leads to Landau quantization 
of the energy levels of the proton in the nuclear medium.
The effects of magnetic field 
and isospin asymmetry on the charmonium decay widths
to $D\bar D$ are found to be quite prominent.
The effects of the anomalous magnetic moments have also been
taken into consideration for obtaining the in-medium masses
of these heavy flavour mesons,
used to calculate the partial decay widths of the 
charmonium states. The medium modifications of the
charmonium decay widths can have observable consequences 
on the production of the charmed mesons in high energy 
asymmetric heavy ion collision experiments.
\end{abstract}

\maketitle

\def\bfm#1{\mbox{\boldmath $#1$}}

\section{Introduction}

The topic of in-medium properties of hadrons and more recently,
of heavy flavour  hadrons \cite{Hosaka_Prog_Part_Nucl_Phys}, 
is an important and challenging area of research
in strong interaction physics, due to its relevance in the 
ultra relativistic heavy ion collision experiments.
The estimated magnetic fields produced
in noncentral ultra high energy nuclear collisions
are huge 
%%%added below the values of the estimates at HIC expts
($eB \sim 2 m_\pi^2$ at RHIC, BNL, and, $eB \sim 15 m_\pi^2$ 
at LHC, CERN) 
%%%added above the values of the estimates at HIC expts
\cite{HIC_mag_1,HIC_mag_2,HIC_mag_3,HIC_mag_4}. This has initiated
studies of the effects of strong magnetic fields on the
in-medium properties of hadrons in the recent years.
However, the time evolution of the magnetic field
\cite{time_evolution_B_HIC_Tuchin_1,time_evolution_B_HIC_Tuchin_2,time_evolution_B_HIC_Tuchin_3,time_evolution_B_HIC_Tuchin_4,time_evolution_B_HIC_Ajit}
is still under debate, which needs the proper estimate
of the electrical conductivity of the medium as well as 
solutions of the magnetohydrodynamic equations.

The properties of the heavy flavour mesons, e.g., 
the open charm and open bottom mesons
\cite{Gubler_D_mag_QSR,machado_1,B_mag_QSR,dmeson_mag,bmeson_mag}
as well as the charmonium states
\cite{charmonium_mag_QSR,charmonium_mag_lee,charmonium_mag,jpsi_etac_mag} 
have been studied in the literature 
in the presence of strong magnetic fields. 
%%%%%%added below in response to comment 3 of Ref. 1 and 2
In the presence of a strong magnetic field, there can be
mixing of the longitudinal components of the vector charmonium states,
e.g, $J/\psi$, $\psi' (\equiv \psi(3686))$, with their pseudoscalar partners
$\eta_c$, $\eta'_c$ \cite{Suzuki_Lee_2017,Alford_Strickland_2013}
which might show in the dilepton spectra as anolamous 
$\eta_c$ and $\eta'_c$ peaks, in addition to the 
$J/\psi$ and $\psi'$ peaks. In Ref. \cite{Suzuki_Lee_2017},
the formation times for the quarkonia have been studied
and it is observed that with increasing magnetic field
strength, the formation times of the vector quarkonia 
become larger, whereas the formation time of their pesudoscalar 
partners become faster \cite{Suzuki_Lee_2017}.
With a larger formation time, the $J/\psi$ could survive 
through the initial thermal effects, thus enhancing its
survivability, whereas the $\eta_c$, $\eta'_c$ 
peaks in the dilepton spectra could probe
an early magnetic field.
%%%%%%added above in response to comment 3 of Ref. 1 and 2
The heavy ion collision experiments involve heavy nuclei
which have much larger number of neutrons than protons,
and hence the isospin asymmetry effects 
on the hadron properties are important to study.
In the present work, we study in-medium decay widths
of the charmonium states ($J/\psi$, $\psi(3686)$, 
$\psi (3770)$, $\chi_{c0}$ and $\chi_{c2}$) to $D\bar D$,
in isospin asymmetric nuclear matter in the presence 
of strong magnetic fields. These decay widths  are computed 
using the $^3P_0$ model \cite{3p0_1,3p0_2,3p0_3,3p0_4,friman,amarvepja}, 
from the mass modifications of the charmonium states
as well as $D$ and $\bar D$ mesons 
calculated using a chiral effective model. 

The heavy quarkonium states, e.g., the charmonium and bottomonium
states have been studied extensively in the literature,
using potential models 
\cite{eichten_1,eichten_2,satz_1,satz_2,satz_3,satz_4,satz_5,repko}.
The effects of the gluonic fluctutations on the  
quarkonium states have been studied in
Refs. \cite{pes1,pes2,voloshin}, with a
color Coulomb potential for the interaction
of the heavy quark, $Q$ and heavy antiquark, $\bar Q$,
within the quarkonium state. Assuming the separation 
of $Q$ and $\bar Q$,
to be small compared to the characteristic scale 
of the gluonic fluctuations, the leading order
contribution of a multipole expansion of the
quarkonium state to the gluonic field leads
to the mass of the quarkonium state
to be proportional to the gluon condensate.
In Ref. \cite{leeko}, the in-medium masses of the charmonium states 
have been studied using leading order QCD formula \cite{pes1} 
and the linear density approximation for
the gluon condensate in the nuclear medium. 
Within the QCD sum rule approach, 
the mass modifications of the charmonium states
are due to the medium changes of the gluon condensate
\cite{kimlee,klingl,amarvjpsi_qsr,jpsi_etac_mag,moritalee_1,moritalee_2,moritalee_3,moritalee_4}, 
whereas, the open heavy flavour mesons are modified due to the
interaction with the light quark condensates in the hadronic
medium 
\cite{open_heavy_flavour_qsr_1,open_heavy_flavour_qsr_2,open_heavy_flavour_qsr_3,open_heavy_flavour_qsr_4,Wang_heavy_mesons_1,Wang_heavy_mesons_2,arvind_heavy_mesons_QSR_1,arvind_heavy_mesons_QSR_2,arvind_heavy_mesons_QSR_3}. 
The heavy flavour mesons have also been studied in the literature,
using quark meson coupling model 
\cite{open_heavy_flavour_qmc_1,open_heavy_flavour_qmc_2,open_heavy_flavour_qmc_3,qmc_1,qmc_2,qmc_3,qmc_4,krein_jpsi,krein_17},
using heavy quark symmetry and interaction
of these mesons with nucleons via pion exchange \cite{Yasui_Sudoh_pion},
using heavy meson effective theory
\cite{Yasui_Sudoh_heavy_meson_Eff_th}, studying the heavy flavour meson as
an impurity in nuclear matter \cite{Yasui_Sudoh_heavy_particle_impurity}
as well as using the coupled channel approach 
\cite{ltolos,ljhs,mizutani_1,mizutani_2,HL,tolos_heavy_mesons_1,tolos_heavy_mesons_2}. 
Within the chiral effective model \cite{Schechter,paper3,kristof1},
the in-medium charmonium masses are
obtained from the medium changes of a scalar dilaton field,
which mimicks the gluon condensates of QCD 
\cite{amarindamprc,amarvdmesonTprc,amarvepja}
and the mass modifications of the $D$ and $\bar D$ mesons
arise due to their interactions with the baryons
and scalar mesons in the hadronic medium
\cite{amarindamprc,amarvdmesonTprc,amarvepja,amdmeson,DP_AM_Ds}.
The chiral effective model has been used extensively in the literature,
for the study of finite nuclei \cite{paper3},
strange hadronic matter \cite{kristof1}, 
light vector mesons \cite{hartree}, 
strange pseudoscalar mesons, e.g. the kaons and antikaons
\cite{kaon_antikaon,isoamss,isoamss1,isoamss2}
in isospin asymmetric hadronic matter,
as well as for the study of bulk matter of neutron stars 
\cite{pneutronstar}.
The light vector mesons ($\omega$, $\rho$
and $\phi$) are modified in the hadronic medium, predominantly 
due to the medium changes in the light 
quark condensates \cite{hatsuda} within 
the QCD sum rule approach. These vector mesons 
in (magnetized) hadronic matter have been 
studied using QCD sum rule calculations from the
medium changes of the light quark condensates
and gluon condensates calculated within the
chiral SU(3) model \cite{am_vecmeson_qsr,vecqsr_mag}.
The kaons and antikaons have been recently studied
in the presence of strong magnetic fields
using this model \cite{kmeson_mag}.
The model has been used to study the open heavy flavour
(charm and bottom) mesons
\cite{amarindamprc,amarvdmesonTprc,amarvepja,amdmeson,DP_AM_Ds,DP_AM_bbar,DP_AM_Bs}, 
the heavy quarkonium states 
\cite{amarvdmesonTprc,amarvepja,AM_DP_upsilon}, the partial decay widths
of the heavy quarkonium states to the open heavy flavour mesons,
in the hadronic medium
\cite{amarvepja} using a light quark creation model \cite{friman},
namely the $^3P_0$ model \cite{3p0_1,3p0_2,3p0_3,3p0_4} as well as 
using a field theoretic model for composite hadrons 
\cite{amspmwg,amspm_upsilon}.
Recently, the effects of magnetic fields on these open and hidden
heavy flavour mesons have been investigated
\cite{dmeson_mag,bmeson_mag,charmonium_mag,upsilon_mag}
using the chiral effective model.

The outline of the paper is as follows : In section II, we describe
briefly the quark pair creation model, namely the $^3P_0$ model
\cite{3p0_1,3p0_2,3p0_3,3p0_4,friman}
used to compute the in-medium partial decay widths of the
charmonium states to $D\bar D$ in magnetized nuclear matter.
The medium modifications of these decay widths are
computed from the mass modifications of the charmonium 
states, $D$ and $\bar D$ mesons calculated within a 
chiral effective model.
In section III, we discuss the results obtained in the present
investigation of these  in-medium charmonium  decay widths
in asymmetric nuclear matter in presence of strong magnetic fields.
In section IV, we summarize the findings of the present study.

\section{Decay widths of Charmonia to $D\bar D$ within $^3P_0$ model} 

In the present work, we compute the in-medium partial decay widths
of the charmonium states ($J/\psi$,
$\psi(3686)$, $\psi(3770)$, $\chi_{c0}$, $\chi_{c2}$) 
to $D\bar D$ ($D^+D^-$ or $D^0\bar {D^0}$)
in magnetized nuclear matter, using the $^3P_0$ model. 
In this  model, a light quark-antiquark pair is created in the 
$^3P_0$  state, and this light quark (antiquark) combines 
with the heavy charm antiquark (charm quark) of the decaying 
charmonium state at rest, resulting in the production of 
the open charm $D$ and $\bar D$ mesons. 

%The general expression 
%for in medium partial width of charmonia is \cite{13}
%\begin{eqnarray}
%\Gamma_{\psi \rightarrow  D{\bar D}} 
%= 2\pi \frac{p_D E_D E_{\bar D}}{ M_{\psi}} \left|M\right|^2 
%\end{eqnarray}
%where M is the invariant matrix element representing 
%the decay of the parent charmonia to  $D\bar{D}$ pairs. 
%This matrix element will
%involve an overlap integral consisting of the momentum 
%space wave functions of the participating mesons, 
%represented by polynomials multiplied by gaussians.
%There are two possible $D {\bar D}$ channels and the various 
%charmonium states can decay through $D^0$ $\bar{D^{0}}$ 
%channel as well as through $D^+$ $D^-$ depending 
%on the masses of the parent and daughter mesons in the medium. 
%The method of calculating the in medium mass modifications 
%of these D mesons and charmonia have been already discussed 
%in the paper\cite{15} where we made use of the chiral 
%effective model. 
The in-medium partial decay widths of charmonium states $\psi(3686)$, 
$\psi(3770)$, $\chi_{c0}$, $\chi_{c2}$ to $D\bar D$ were studied
using $^3 P_0$ model in Ref. \cite{friman}, where the masses 
of the $D$ and $\bar D$ mesons 
were assumed to have same medium modification, in addition
to the degeneracy of the masses of the $D^+$ and $D^0$ 
within the $D$ doublet, and of $D^-$ and $\bar {D^0}$ 
within the $\bar D$ doublet in symmetric nuclear matter.
In the asymmetric strange hadronic matter,
the partial decay widths of the charmonium states,
$J/\psi$, $\psi(3686)$ and $\psi(3770)$ 
to $D\bar D$ pair 
were calculated within the $^3P_0$ model,
using the mass modifications of these charmonium 
states, $D$ and $\bar D$ mesons calculated within
the chiral effective model \cite{amarvepja}. 

%**********************************************
%*****Discussions below regd D and D bar masses
%in response to comment 1 of Referee 1 and 2 *****
%**************************************************
The mass modifications of the open charm meson 
are calculated from their interactions with the
isoscalar scalar mesons ($\sigma$ and $\zeta$),
isovector scalar meson, $\delta$,
and the nucleons in the magnetized asymmetric nuclear matter
\cite{dmeson_mag}, 
whereas the masses of the charmonium states 
are modified due to the medium change of the gluon
condensates, simulated by the scalar dilation field,
$\chi$ \cite{charmonium_mag}  within the chiral effective model.
The in-medium masses of the $D$ and $\bar D$ mesons
in the magnetied asymmetric nuclear matter
were calculated in Ref. \cite{dmeson_mag},
within the frozen glueball approximation,
i.e., neglecting the medium modifications of the dilaton
field, $\chi$.
In the present work, we take into account the
medium modification of the dilaton field, $\chi$, 
by solving the Euler Lagrange equations of motions
for the scalar fields, $\sigma$, $\zeta$ and $\delta$, 
as well as $\chi$ self consistently,
for given values of baryon density, isospin asymmetry
parameter and magnetic field. Accounting for the medium
modifications of the $\chi$ field, however, are observed to give rise 
to marginal modifications to the masses of the $D$ and $\bar D$
meson masses as compared to the case of frozen glueball
approximation.
The medium modifications of the charmonium decay widths 
to $D\bar D$ in the magnetized nuclear matter arise
due to the mass modifications of the open charm and charmonium states,
which depend on the medium changes of the scalar fields and the dilaton
field. To undertand the in-medium behaviour of these fields,
which modify the masses of the charmonium and open charm mesons, 
and hence the charmonium decay widths to $D\bar D$,
we write explicitly the coupled equations of motion of the scalar fields 
($\sigma$, $\zeta$, $\delta$) and the dilaton field, $\chi$. 
These are given as \cite{upsilon_mag}
\begin{eqnarray}
&& k_{0}\chi^{2}\sigma-4k_{1}\left( \sigma^{2}+\zeta^{2}
+\delta^{2}\right)\sigma-2k_{2}\left( \sigma^{3}+3\sigma\delta^{2}\right)
- 2k_{3}\chi\sigma\zeta 
-\frac{d}{3} \chi^{4} \bigg (\frac{2\sigma}{\sigma^{2}-\delta^{2}}\bigg )
\nonumber \\
&+&\left( \frac{\chi}{\chi_{0}}\right) ^{2}m_{\pi}^{2}f_{\pi}
-\sum g_{\sigma i}\rho_i^s = 0 
\label{sigma}
\end{eqnarray}
\begin{eqnarray}
&& k_{0}\chi^{2}\zeta-4k_{1}\left( \sigma^{2}+\zeta^{2}+\delta^{2}\right)
\zeta-4k_{2}\zeta^{3}-k_{3}\chi\left( \sigma^{2}-\delta^{2}\right)
-\frac{d}{3}\frac{\chi^{4}}{\zeta}
\nonumber\\
&+&\left(\frac{\chi}{\chi_{0}} \right) 
^{2}\left[ \sqrt{2}m_{k}^{2}f_{k}-\frac{1}{\sqrt{2}} m_{\pi}^{2}f_{\pi}\right]
 - \sum g_{\zeta i}\rho_i^s = 0 
\label{zeta}
\end{eqnarray}
\begin{eqnarray}
& & k_{0}\chi^{2}\delta-4k_{1}\left( \sigma^{2}+\zeta^{2}+\delta^{2}\right)
\delta-2k_{2}\left( \delta^{3}+3\sigma^{2}\delta\right) +2k_{3}\chi\delta 
\zeta 
\nonumber\\
& +&   \frac{2}{3} d \chi^{4} \left( \frac{\delta}{\sigma^{2}-\delta^{2}}\right)
-\sum g_{\delta i}\rho_i^s = 0
\label{delta}
\end{eqnarray}
\begin{eqnarray}
& & k_{0}\chi \left( \sigma^{2}+\zeta^{2}+\delta^{2}\right)-k_{3}
\left( \sigma^{2}-\delta^{2}\right)\zeta + \chi^{3}\left[1
+{\rm {ln}}\left( \frac{\chi^{4}}{\chi_{0}^{4}}\right)  \right]
%+(4k_{4}-d)\chi^{3}
+ 4k_{4}\chi^{3}
\nonumber\\
& -&  \frac{4}{3} d \chi^{3} {\rm {ln}} \Bigg ( \bigg (\frac{\left( \sigma^{2}
-\delta^{2}\right) \zeta}{\sigma_{0}^{2}\zeta_{0}} \bigg ) 
\bigg (\frac{\chi}{\chi_0}\bigg)^3 \Bigg )
+\frac{2\chi}{\chi_{0}^{2}}\left[ m_{\pi}^{2}
f_{\pi}\sigma 
 +\left(\sqrt{2}m_{k}^{2}f_{k}-\frac{1}{\sqrt{2}}
m_{\pi}^{2}f_{\pi} \right) \zeta\right]  = 0 
\label{chi}
\end{eqnarray}
In the above, $\rho_i^s (i=p,n)$ are the scalar densities 
for the nucleons, and, $\sigma_0$, $\zeta_0$ and $\chi_0$
are the vacuum values for these scalar fields.
%**********************************************
%*****Discussions below regd D and D bar masses
%in response to comment 1 of Referee 1 and 2 **********
%**************************************************

The expressions for the decay widths of the charmonium states 
in the asymmetric nuclear matter, considered in the present 
investigation, are given as \cite{friman}
\begin{eqnarray}
\Gamma_{J/\psi \rightarrow  D{\bar D}} & =& 
\frac { \sqrt {\pi}{E_D E_{\bar D}}\gamma^2}{2 M_{J/\psi}}
\frac{2^8 r^3(1+r^2)^2}{3(1+2r^2)^5} x^3 
%\nonumber \\
\times \exp\Bigg(-\frac{x^2}{2(1+2r^2)}\Bigg), 
\label{jpsidw}
 \end{eqnarray}
 \begin{eqnarray}
  \Gamma_{\psi(3686) \rightarrow  D {\bar D}} 
&= & \frac{\sqrt {\pi}{E_D E_{\bar D}}\gamma^2}
{2 M_{\psi(3686)}}\frac{2^7(3+2r^2)^2 
(1-3r^2)^2}{3^2(1+2r^2)^7} x^3\nonumber \\
&\times & \Bigg(1+
\frac{2r^2(1+r^2)}{(1+2r^2)(3+2r^2)(1-3r^2)}x^2\Bigg)^2 
\times \exp \Bigg ({-\frac{x^2}{2(1+2r^2)}}\Bigg),
\label{psipdw}
 \end{eqnarray}
  \begin{eqnarray}
  \Gamma_{\psi(3770) \rightarrow  D{\bar D}} 
&=& \frac{\sqrt {\pi}{E_D E_{\bar D}}\gamma^2}{2 M_{\psi(3770)}}
\frac{2^{11} 5}{3^2}\Bigg(\frac{r}{1+2r^2}\Bigg)^7 
\nonumber \\ &\times & x^3
\Bigg(1-\frac{1+r^2}{5(1+2r^2)}x^2\Bigg)^2 
\times  
 \exp \Bigg ({-\frac{x^2}{2(1+2r^2)}}\Bigg),
\label{psippdw}
 \end{eqnarray}
 \begin{eqnarray}
  \Gamma_{\chi_{c0} \rightarrow  D{\bar D}} 
&=& \frac{ \sqrt {\pi}E_D E_{\bar {D}}\gamma^2}{2 M_{\chi_{c0}}}2^{9} 
3\Bigg(\frac{r}{1+2r^2}\Bigg)^5 
 \nonumber \\ &\times &  
 x \Bigg(1-\frac{1+r^2}{3(1+2r^2)}x^2\Bigg)^2 
\times 
\exp \Bigg({-\frac{x^2}{2(1+2r^2)}}\Bigg),
\label{chic0dw}
 \end{eqnarray}
  \begin{eqnarray}
  \Gamma_{\chi_{c2} \rightarrow  D\bar{D}} 
&= & \frac{\sqrt {\pi} E_D E_{\bar {D}}\gamma^2}
{2 M_{\chi_{c2}}}\frac{2^{10}}{15} 
\frac{r^5(1+r^2)^2}{(1+2r^2)^7} 
\times x^5 \exp \Bigg({-\frac{x^2}{2(1+2r^2)}}\Bigg).
\label{chic2dw}
 \end{eqnarray}
In the above, $M_{\Psi}$ is the in-medium mass of the corresponding 
charmonium state 
($\Psi=J/\psi,\psi(3686),\psi(3770),\chi_{c0},\chi_{c2}$), 
$E_{D}$, $E_{\bar D}$ are energies of the outgoing $D$
and ${\bar D}$ mesons given as
\begin{eqnarray}
E_D = (p_D^2 + m_D^2)^{1/2},\;\;
E_{\bar D} = (p_D^2 + m_{\bar D}^2)^{1/2},
\label{ededbar}
\end{eqnarray}
with $m_D $ and $m_{\bar D}$ as the in-medium 
masses of the $D$ and $\bar D$ mesons, and  
$p_D$ is the 3-momentum of the produced open charm mesons 
in the centre of mass frame given as
\begin{eqnarray}
p_D= \Bigg(\frac{ M_{\psi}^2}{4}- \frac{ m_D^2
 + m_{\bar D}^2}{2}
+\frac{( m_D^2 - m_{\bar D}^2)^2}{4M_{\psi}^2}\Bigg) ^{1/2}.
\label{pd}
\end{eqnarray}
In the expressions for the partial decay widths of the 
charmonium states to $D\bar D$,  $\gamma$ is the coupling 
strength related to  the strength of the $^3P_0$ vertex. 
It signifies the probability for creating a light quark-antiquark pair. 
The wave functions of the charmonium states as well as
$D(\bar D)$ mesons are assumed to be the wave functions
with harmonic oscillator potential.
The ratio $r=\beta/\beta_D$,
where $\beta$ is the strength of the harmonic 
potential of the parent chamonium state and $\beta_D$ 
is the strength of harmonic potential of the daughter 
$D(\bar D)$-meson. The scaled momentum $x$ is defined 
as $ x=p_D/\beta_D $. In this study the value 
of $\gamma$ is chosen to be 0.33 which reproduces the
observed decay widths of $\psi(3770)$ to $D^+D^-$
as well as to $D^0 \bar {D^0}$ in vacuum \cite{amarvepja}.
The value of $\beta_D$ is taken as 0.31 GeV, consistent 
with the decay widths of $\psi(4040)$ to $D\bar D$,
$D^* \bar D$, $D \bar {D^*}$ and $D^* \bar {D^*}$ 
in vaccum \cite{leeko}.
%%%added below discussion about the behaviour of 
%%%% decay width as density
As we might see from the expressions of the
decay widths of the charmonium states 
($J/\psi, \psi(3686),\psi(3770), \chi_{c0}, \chi_{c2}$)
to $D\bar D$ given by the equations  
(\ref{jpsidw})--(\ref{chic2dw}), these decay widths 
depend on the variable,
$x$, which is the center of mass momentum, $p_D$ in units 
of $\beta_D$, as a polynomial multiplied by an exponential
term. The dependence of these decay widths on the density
at given values of isospin asymmetry and magnetic field
are thus determined by their dependence on $p_D$, given by
equation (\ref{pd}), in terms of the in-medium masses
of the charmonium state as well as $D$ and $\bar D$ mesons.
Depending on the form of the polynomial, 
the in-medium decay widths of these charmonium states,
are observed to have very different behaviour.
%%%added below discussion about the behaviour of 
%%%% decay width as density

We use an effective chiral $SU(3)$ model \cite{paper3} 
to study the in-medium masses
of the charmonium states in asymmetric nuclear matter
in the presence of an external magnetic field \cite{charmonium_mag}.
The model is based on the nonlinear realization of chiral 
symmetry \cite{weinberg,coleman,bardeen} and broken scale invariance 
\cite{paper3,kristof1,hartree}. 
The scale invariance breaking is through a logarithmic 
potential given in terms of a scalar dilaton field \cite{heide1},
and the medium modification of the dilaton field gives 
the medium modification of the gluon condensate,
used to calculate the charmonium mass in the nuclear
medium in the presence of an external magnetic field. 
The contribution of the magnetic field
is incorporated \cite{broderick1,broderick2,Wei,mao}
into the chiral effective model
to study the mass modifications of the charmonium states 
\cite{charmonium_mag}
as well as open charm mesons \cite{dmeson_mag} 
in the magnetized nuclear matter, including the effects
of anomalous magnetic moments (AMM) of the nucleons
\cite{broderick1,broderick2,Wei,mao,amm,VD_SS_1,VD_SS_2,aguirre_fermion}.
%%%%Comment regd comment 2 of Referee 1 added below
The AMM effects are taken into consideration in the present 
study of the partial
charmonium decay widths to $D\bar D$ in the magnetized nuclear
matter, through the in-medium masses of the charmonium state
as well as the open charm mesons.
In the presence of the AMM effects, there are contributions
to the nucleon fields, due to the tensorial interaction of the 
nucleons with the electromagnetic field 
$(-\frac {1}{4} \kappa_i \mu_N {\bar {\psi_i}} 
\sigma ^{\mu \nu}F_{\mu \nu} \psi_i)$, in addition 
to the vectorial interaction
term $(-{\bar {\psi_i}}q_i \gamma_\mu A^\mu \psi_i)$,
where, $\psi_i$ corresponds to the $i$-th baryon (proton and neutron 
for nuclear matter, as considered in the present work)
\cite{dmeson_mag,bmeson_mag,charmonium_mag,broderick1,broderick2,Wei,mao,amm,VD_SS_1,VD_SS_2,aguirre_fermion}.
The values of $\kappa_p$ and  $\kappa_n$ are given as 
$3.5856$ and $-3.8263$ respectively, which are the values
of the gyromagnetic ratio corresponding to the 
anomalous magnetic moments of the proton and 
neutron respectively.
The scalar densities and number densities
of the nucleons are modified with the AMM effects
due to the tensorial interaction term.
With inclusion of the AMM effects, the values for the scalar fields and
dilaton field, which are obtained by solving the coupled equations
of motion (\ref{sigma})--(\ref{chi}), are modified, as compared
to when the AMM effects are not taken into consideration.
Hence, the mass shifts of the charmonium states, as well as the $D$ and 
$\bar D$ meson are modified, which in turn modify the charmonium 
decay widths to $D\bar D$. 
%%%%Comment regd comment 2 of Referee 1 added above

As has already been mentioned, the mass shift in the charmonium
state arises due to the medium modification of the
scalar gluon condensate, and hence due to the change 
in the value of the dilaton field
in the chiral effective model, and is given as 
\cite{leeko,amarvepja,charmonium_mag}
\begin{equation}
\Delta m_{\Psi}= \frac{4}{81} (1 - d) \int dk^{2} 
\langle \vert \frac{\partial \psi (\vec k)}{\partial {\vec k}} 
\vert^{2} \rangle
%\langle \vert  {\vec {\bigtriangledown} \psi} (\vec k) \vert^{2} \rangle
\frac{k}{k^{2} / m_{c} + \epsilon}  \left( \chi^{4} - {\chi_0}^{4}\right), 
\label{masspsi}
\end{equation}
where 
\begin{equation}
\langle \vert \frac{\partial \psi (\vec k)}{\partial {\vec k}} 
\vert^{2} \rangle
=\frac {1}{4\pi}\int 
\vert \frac{\partial \psi (\vec k)}{\partial {\vec k}} \vert^{2}
d {\rm \Omega}.
\end{equation}
In the above, $m_c$ is the mass of the charm quark, taken as 1.95 GeV,
$m_\Psi$ is the vacuum mass of the charmonium state 
and $\epsilon = 2 m_{c} - m_{\Psi}$ is the binding energy. 
$\psi ({\bf k})$ is the wave function of the charmonium state
in the momentum space, which has been assumed to be Gaussian
\cite{friman,amarvepja,leeko}.

\section{Results and Discussions}

We investigate the in-medium partial decay widths 
of the charmonium state $\psi$ ($J/\psi$, $\psi(3686)$,
$\psi(3770)$, $\chi_{c0}$ and $\chi_{c2}$) to 
$D\bar D$ pair in the magnetized nuclear matter.
The medium modifications of these decay widths
are calculated from the medium modifications
of their masses. The in-medium masses of the
charmonium states, $J/\psi$, $\psi(3686)$ and
$\psi(3770)$ in asymmetric nuclear matter
in the presence of strong magnetic fields,
have already been studied using a chiral effective model
in Ref. \cite{charmonium_mag}.
These masses were calculated from the medium
changes of a scalar dilaton field, which simulates 
the gluon condensates of QCD, within the effective 
hadronic model through broken scale invariance.
In the present work, the mass modifications of the
1P charmonium states, $\chi_{c0}$
as well as $\chi_{c2}$
have been studied, which are obtained from the medium
modifications of the scalar dilaton field, $\chi$,
from the equation (\ref{masspsi}), using the
wave function for the charmonium state to be
of harmonic oscillator wave function for 1P state.  
The values of the strength of the harmonic oscillator 
potential, $\beta$ 
%(in GeV) 
for $J/\psi$, $\psi(3686)$ and $\psi(3770)$
(assuming these states to be 1S, 2S and 1D states)
are calculated 
%to be 0.52, 0.38 and 0.37, 
from their rms radii $\langle r^{2} \rangle$ of 0.47$^{2}$ fm$^2$, 
0.96$^{2}$ fm$^2$ and 1 fm$^{2}$ respectively
\cite{leeko,amarvepja}.
For the 1P states, $\chi_{c0}$ and $\chi_{c2}$, 
we extract the values of $\beta$ from a
linear extrapolation of the vacuum mass versus $\beta$ 
graph of these 1P states as well as the $\psi(3686)$ and
$\psi(3770)$. The values of $\beta$ for $\chi_{c0}$
(vacuum mass of 3414.7 MeV) and $\chi_{c2}$ 
(vacuum mass of 3556.17 MeV) are obtained as
0.44 GeV and 0.41 GeV respectively.

\begin{figure}
\includegraphics[width=16cm,height=16cm]{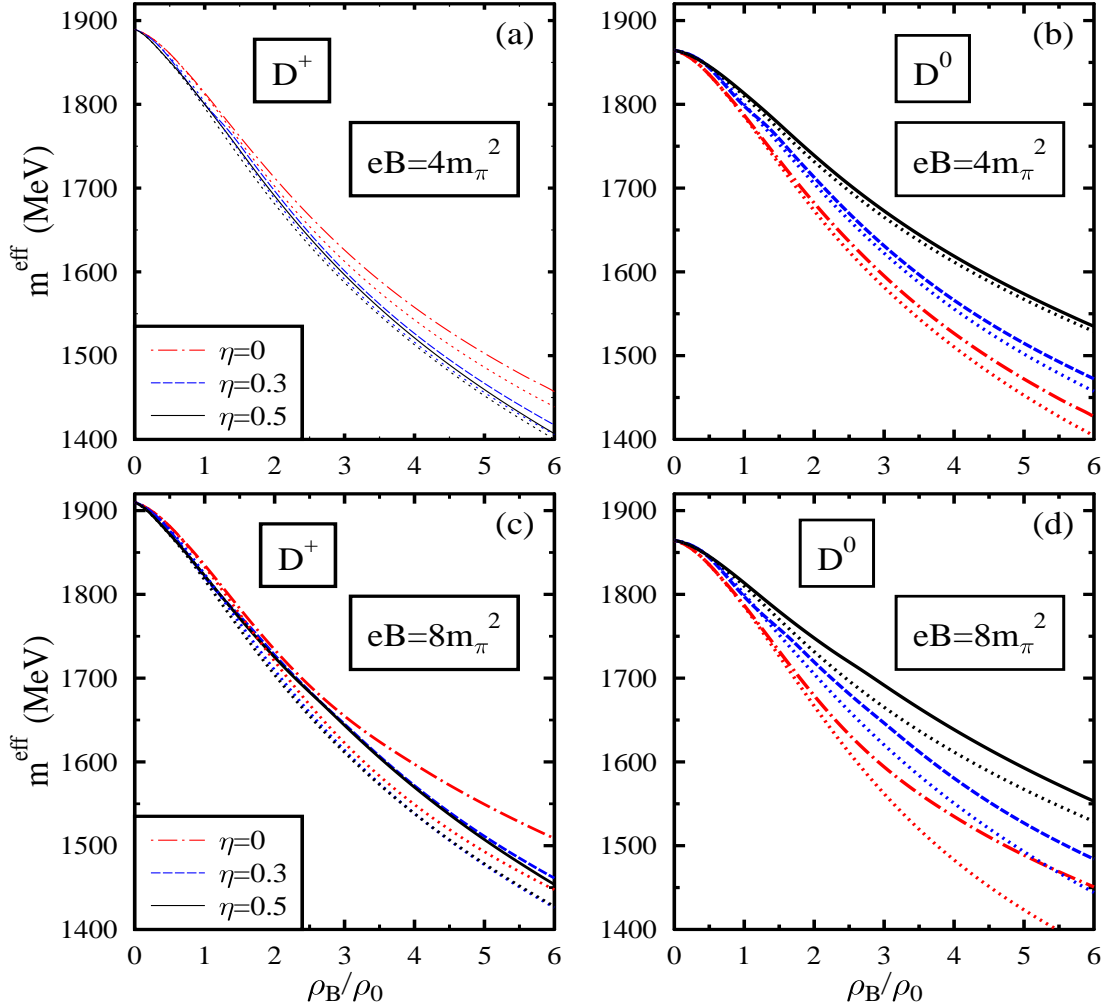}
%\hskip -0.5in
%\includegraphics[width=10cm,height=10cm]{f1.eps}
%\resizebox{0.6\textwidth}{!}{%
%\includegraphics{f1.eps}
%}
\caption{(Color online)
The masses of $D^+$ and $D^0$ mesons plotted as functions of the
baryon density in units of nuclear matter saturation density,
for different values of the magnetic field and isospin 
asymmetry parameter, $\eta$, including the effects of the
anomalous magnetic moments of the nucleons. The results
are compared to the case when the effects of anomalous magnetic 
moments are not taken into consideration (shown as dotted lines).
}
\label{md_chi_mag}
\end{figure}
\begin{figure}
\includegraphics[width=16cm,height=16cm]{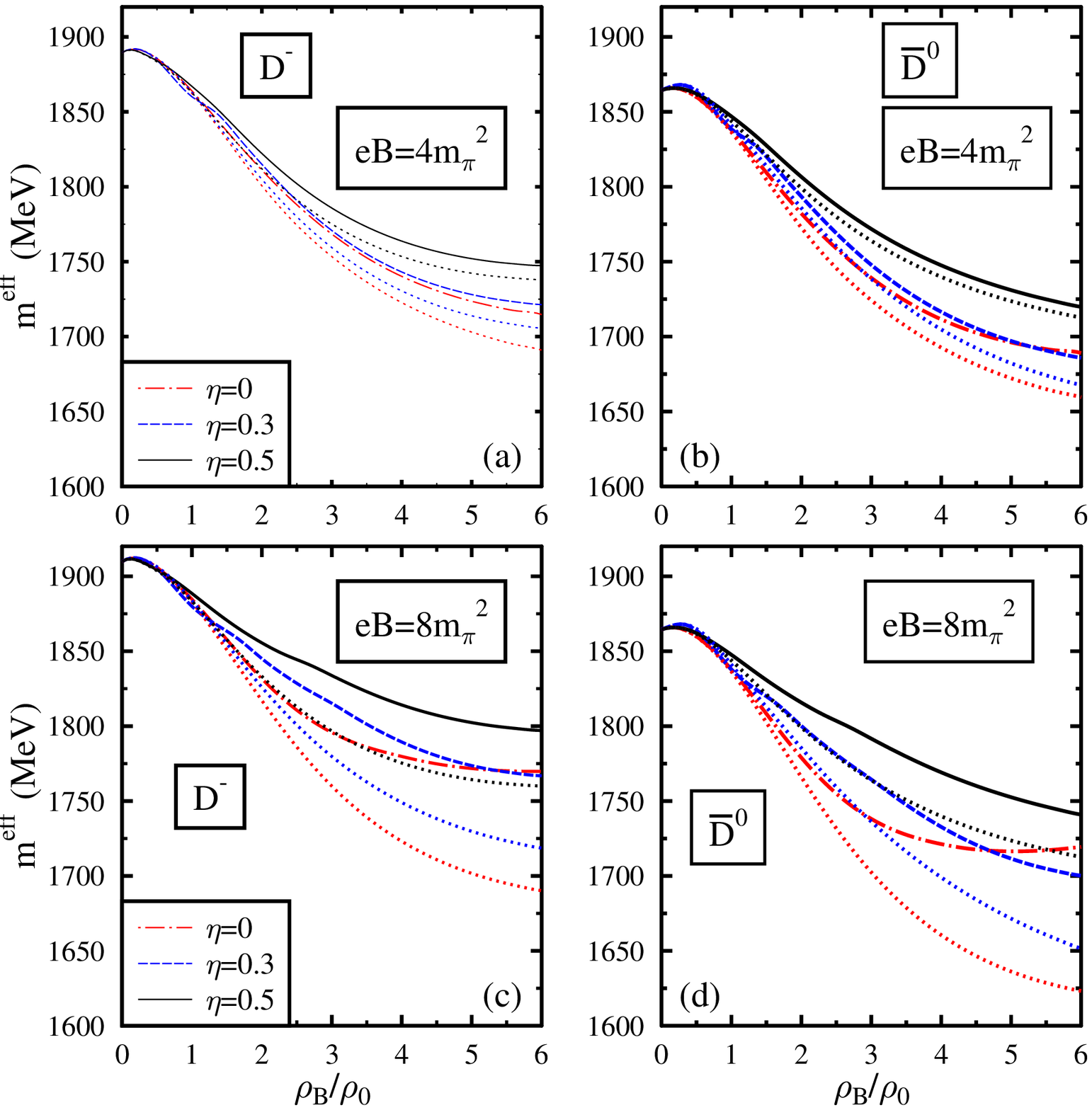}
%\hskip -0.5in
%\includegraphics[width=10cm,height=10cm]{f2.eps}
%\resizebox{0.6\textwidth}{!}{%
%\includegraphics{f1.eps}
%}
\caption{(Color online)
The masses of $D^-$ and $\bar {D^0}$ mesons plotted as functions of the
baryon density in units of nuclear matter saturation density,
for different values of the magnetic field and isospin 
asymmetry parameter, $\eta$, including the effects of the
anomalous magnetic moments of the nucleons. The results
are compared to the case when the effects of anomalous magnetic 
moments are not taken into consideration (shown as dotted lines).
}
\label{mdbar_chi_mag}
\end{figure}
\begin{figure}
\includegraphics[width=16cm,height=16cm]{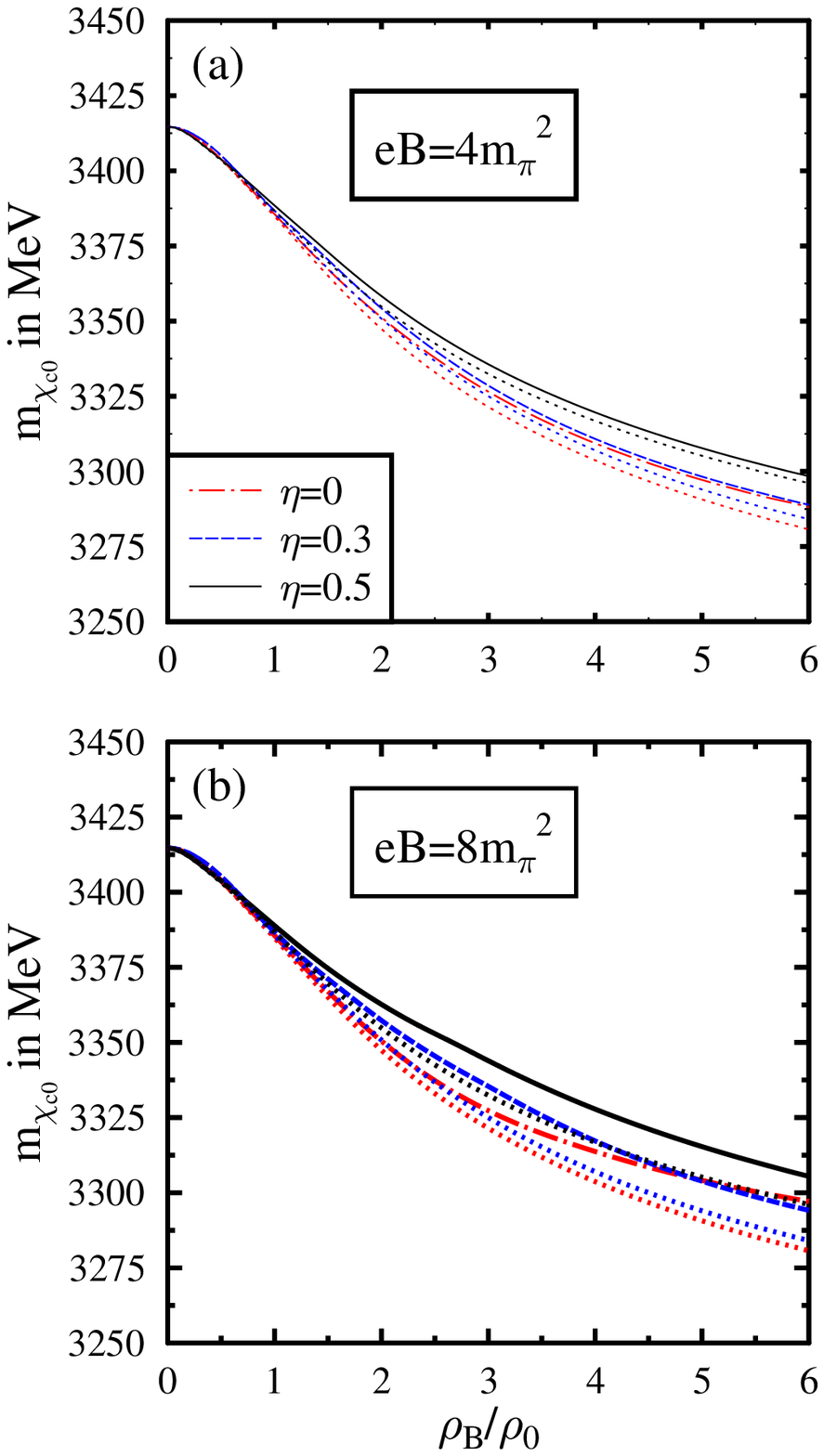}
\hskip -0.5in
%\includegraphics[width=12cm,height=12cm]{f3.eps}
%\resizebox{0.6\textwidth}{!}{%
%\includegraphics{f1.eps}
%}
\caption{(Color online)
The mass of $\chi_{c0}$ is plotted as a function of the
baryon density in units of nuclear matter saturation density,
for different values of the magnetic field and isospin 
asymmetry parameter, $\eta$, including the effects of the
anomalous magnetic moments of the nucleons. The results
are compared to the case when the effects of anomalous magnetic 
moments are not taken into consideration (shown as dotted lines).
}
\label{mchic0_mag}
\end{figure}
\begin{figure}
\hskip -0.5in
\includegraphics[width=16cm,height=16cm]{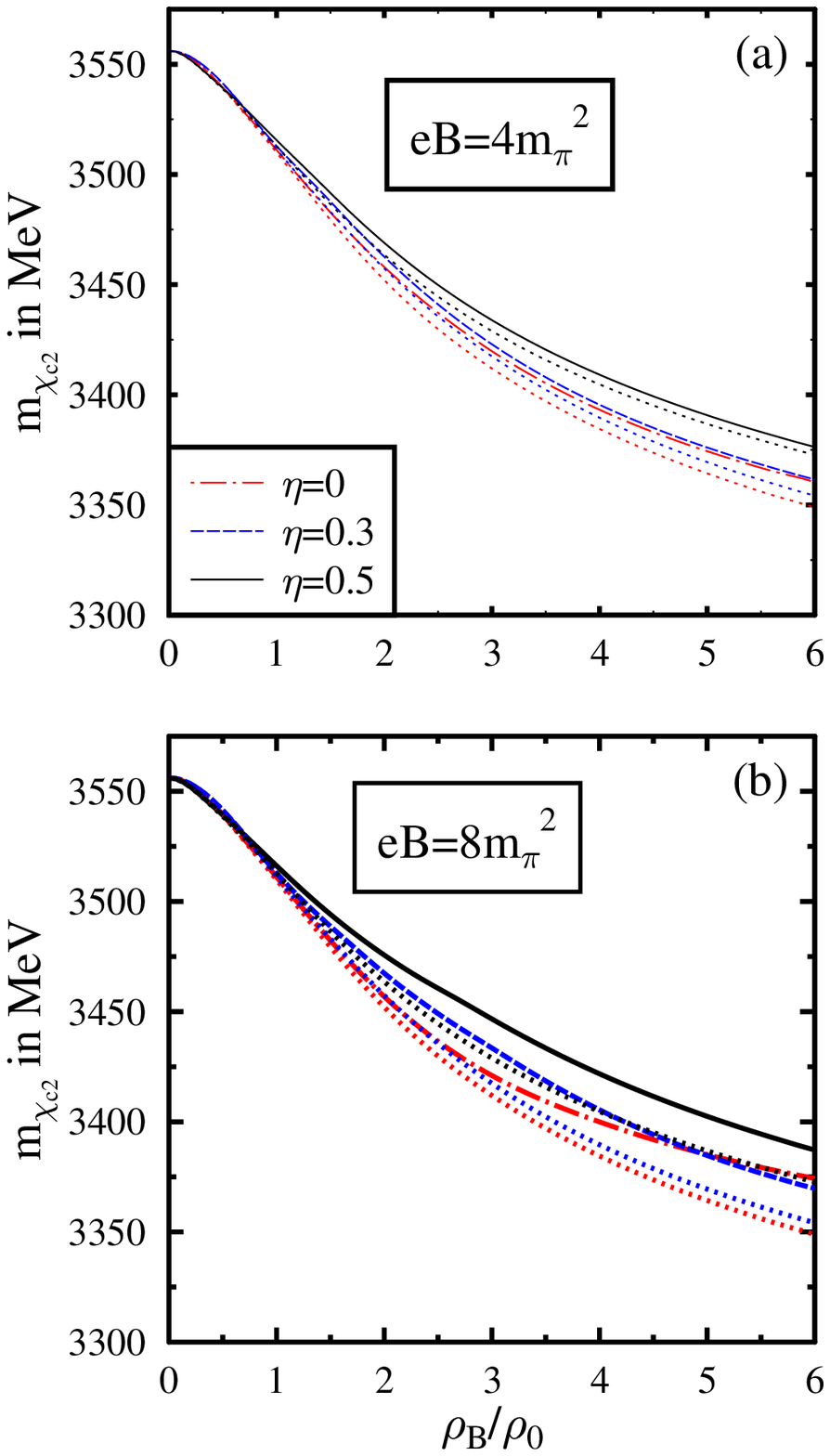}
\caption{(Color online)
The mass of $\chi_{c2}$ is plotted as a function of the
baryon density in units of nuclear matter saturation density,
for different values of the magnetic field and isospin 
asymmetry parameter, $\eta$, including the effects of the
anomalous magnetic moments of the nucleons. The results
are compared to the case when the effects of anomalous magnetic 
moments are not taken into consideration (shown as dotted lines).
}
\label{mchic2_mag}
\end{figure}
\begin{figure}
\hskip -0.3in
\includegraphics[width=16cm,height=16cm]{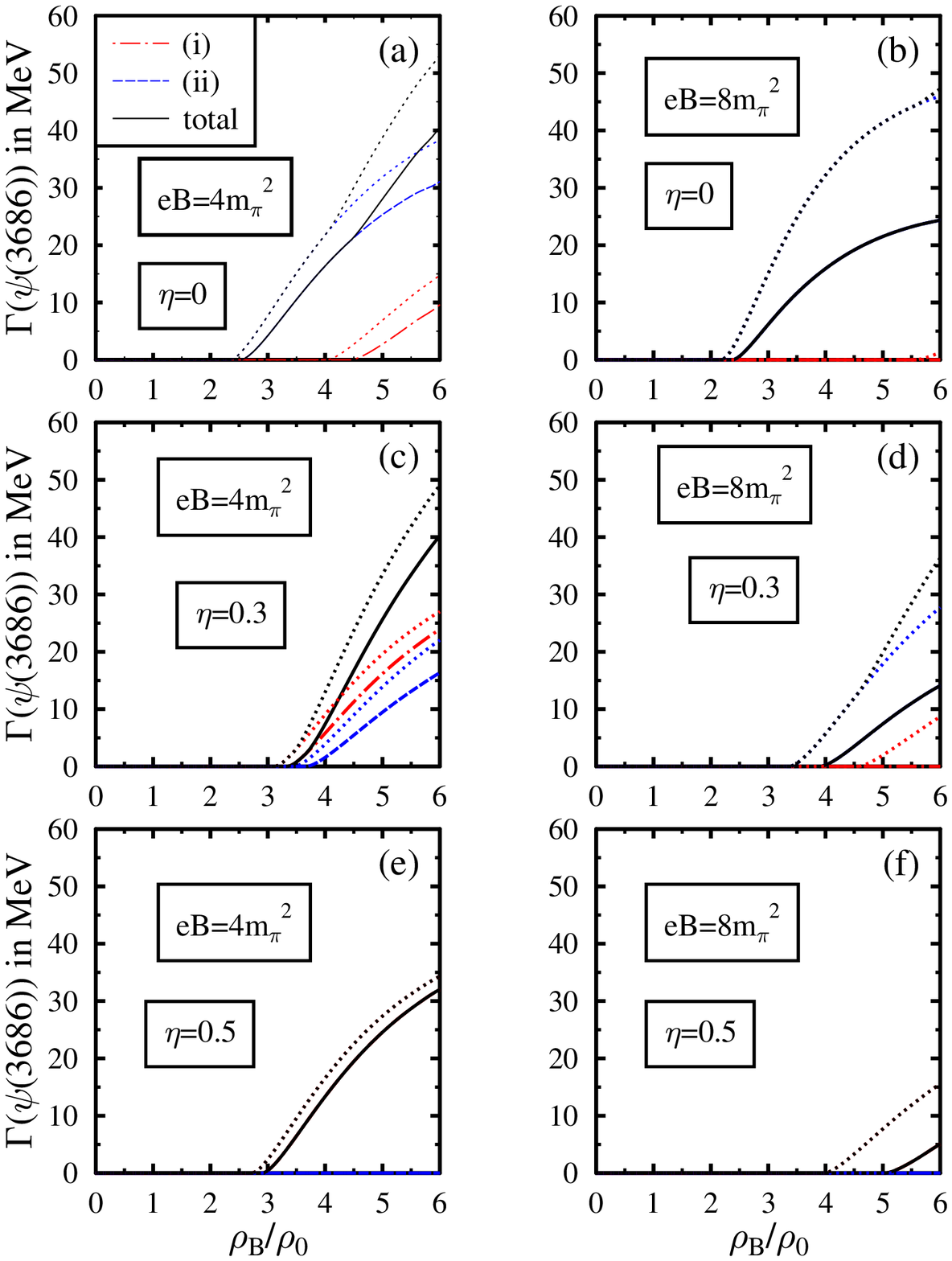}
\caption{(Color online)
Decay widths of $\psi (3686)$ to $D\bar D$ plotted as functions of the
baryon density in units of nuclear matter saturation density,
for the channels (i) $D^+D^-$, (ii) $D^0 \bar {D^0}$
and (iii) total of these two channels.
These are shown for different values of the magnetic field 
and isospin asymmetry parameter, $\eta$, including the effects 
of the anomalous magnetic moments of the nucleons. The results
are compared to the case when the effects of anomalous magnetic 
moments are not taken into consideration (shown as dotted lines).
}
\label{psi3686_decay_mag}

\end{figure}
\begin{figure}
\hskip -0.3in
\includegraphics[width=16cm,height=16cm]{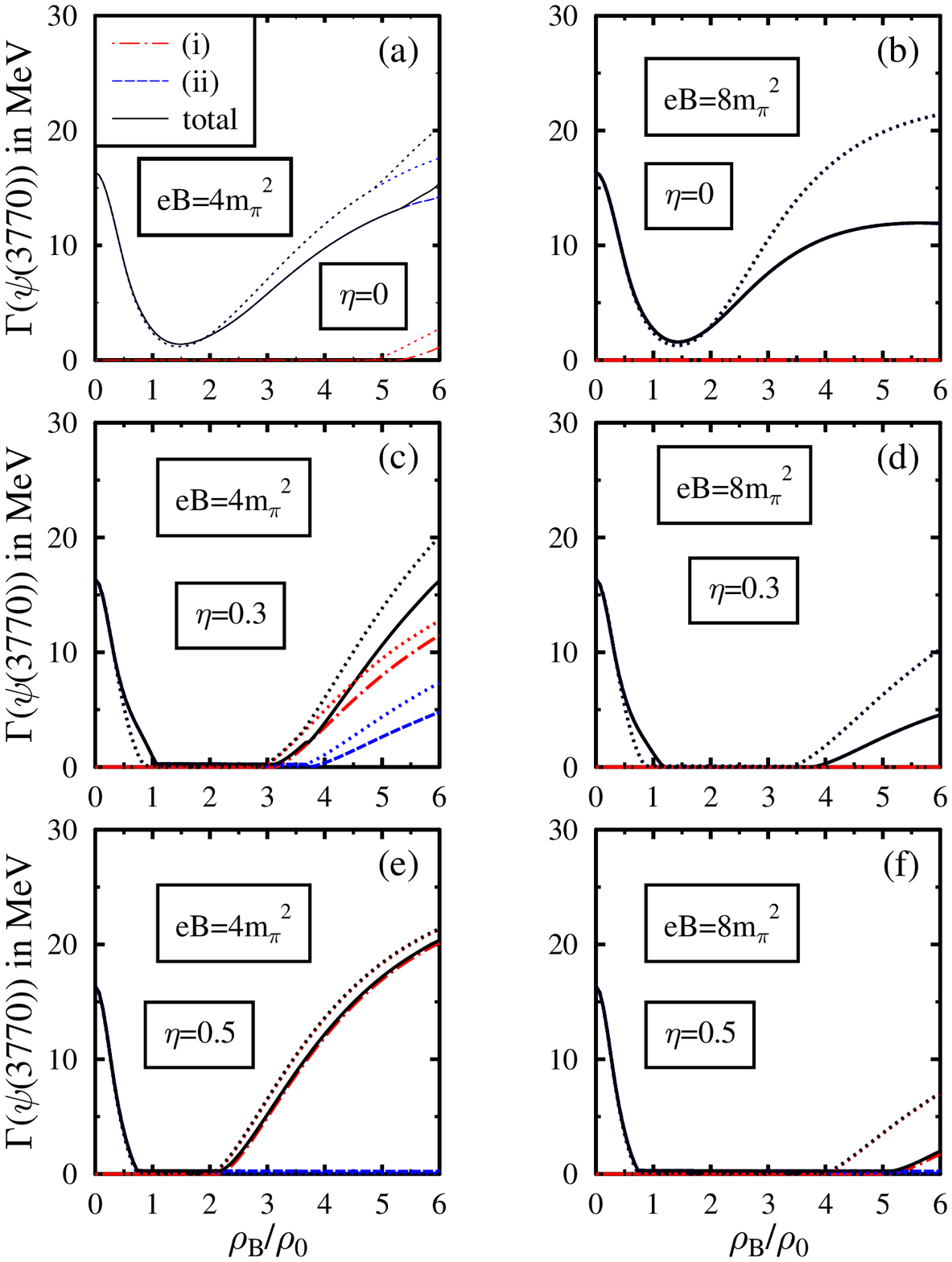}
\caption{(Color online)
Decay widths of $\psi (3770)$ to $D\bar D$ plotted as functions of the
baryon density in units of nuclear matter saturation density,
for the channels (i) $D^+D^-$, (ii) $D^0 \bar {D^0}$
and (iii) total of these two channels.
These are shown for different values of the magnetic field 
and isospin asymmetry parameter, $\eta$, including the effects 
of the anomalous magnetic moments of the nucleons. The results
are compared to the case when the effects of anomalous magnetic 
moments are not taken into consideration (shown as dotted lines).
}
\label{psi3770_decay_mag}
\end{figure}

\begin{figure}
\hskip -0.3in
\includegraphics[width=16cm,height=16cm]{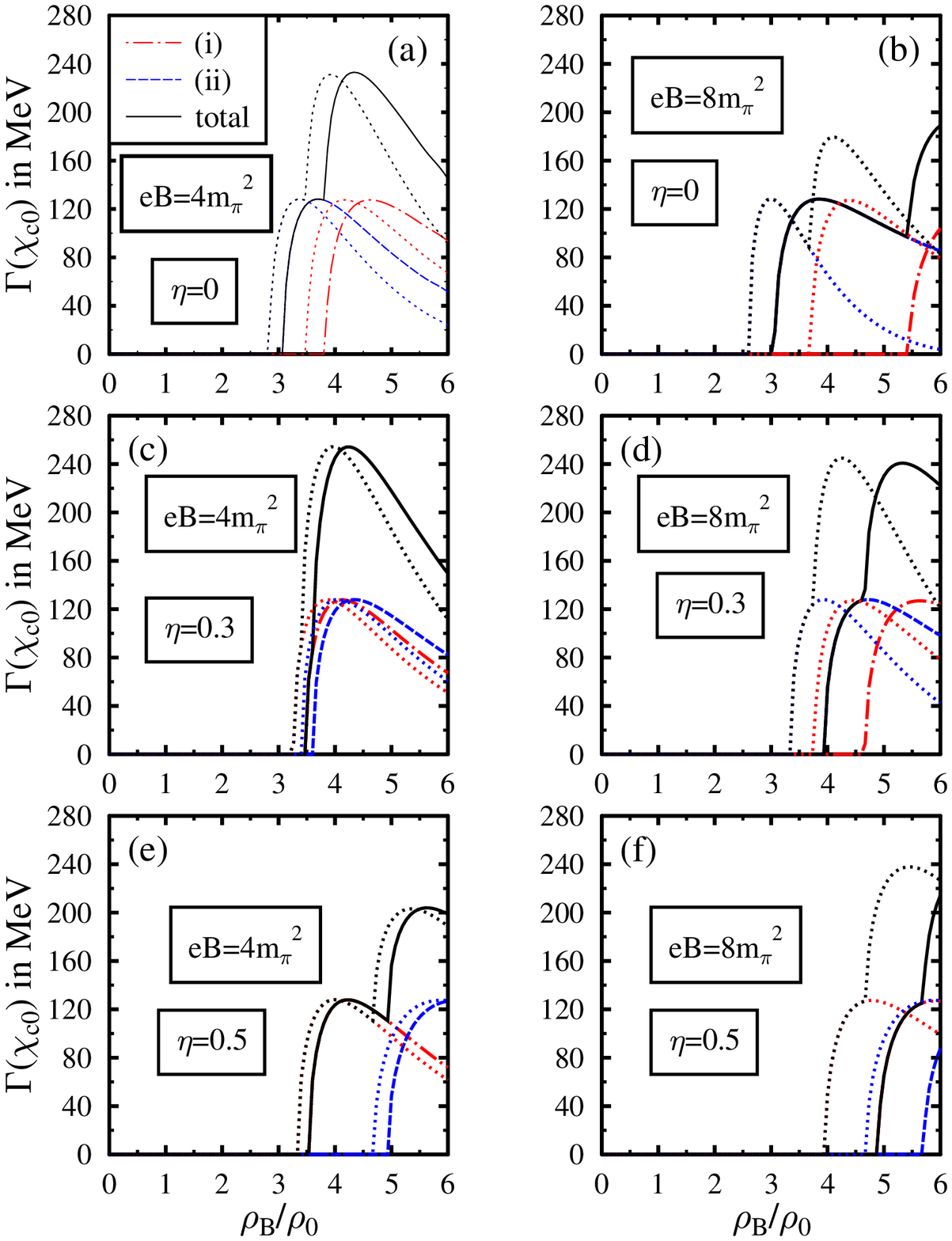}
\caption{(Color online)
Decay widths of $\chi_{c0}$ to $D\bar D$ plotted as functions of the
baryon density in units of nuclear matter saturation density,
for the channels (i) $D^+D^-$, (ii) $D^0 \bar {D^0}$
and (iii) total of these two channels.
These are shown for different values of the magnetic field 
and isospin asymmetry parameter, $\eta$, including the effects 
of the anomalous magnetic moments of the nucleons. The results
are compared to the case when the effects of anomalous magnetic 
moments are not taken into consideration (shown as dotted lines).
}
\label{chic0_decay_mag}

\end{figure}
\begin{figure}
\hskip -0.3in
\includegraphics[width=16cm,height=16cm]{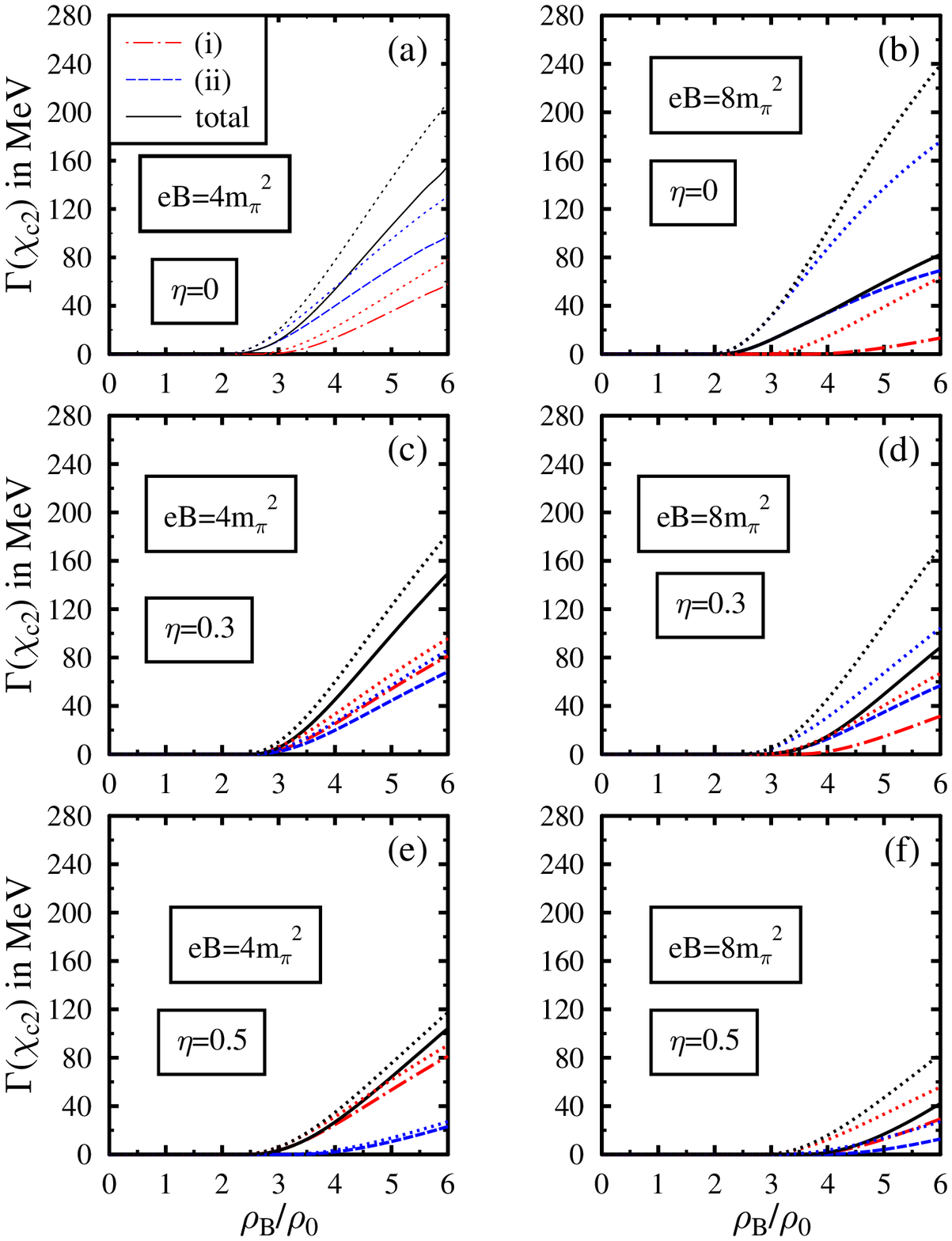}
\caption{(Color online)
Decay widths of $\chi_{c2}$ to $D\bar D$ plotted as functions of the
baryon density in units of nuclear matter saturation density,
for the channels (i) $D^+D^-$, (ii) $D^0 \bar {D^0}$
and (iii) total of these two channels.
These are shown for different values of the magnetic field 
and isospin asymmetry parameter, $\eta$, including the effects 
of the anomalous magnetic moments of the nucleons. The results
are compared to the case when the effects of anomalous magnetic 
moments are not taken into consideration (shown as dotted lines).
}
\label{chic2_decay_mag}
\end{figure}

The masses of the $D$ and $\bar D$ mesons are obtained
in the present work, from their interactions with the
nucleons and the scalar mesons, $\sigma$, $\zeta$
and $\delta$ in the isospin asymmetric magnetized
nuclear matter. The values of these meson fields are solved 
from the coupled equations of these fields as
well as of the scalar dilaton field in the mean field
approximation. 
The study of $D (\bar D)$ mesons
in asymmetric nuclear matter in the presence of
a strong magnetic field in Ref. \cite{dmeson_mag},
used the frozen glueball aprroximation,
i.e., $\chi$ to be fixed at the vacuum value of $\chi$ 
as 409.77 MeV. The masses of the $D$ and $\bar D$ mesons in the
magnetized nuclear matter in the present work are observed
to have modifications (though small) as compared to the masses obtained 
in Ref. \cite{dmeson_mag},
due to the medium dependence of the dilaton field $\chi$ considered 
in the present investigation.

The masses of the $D (D^+,D^0)$ and $\bar D (D^-,\bar {D^0})$ mesons
are shown in figures \ref{md_chi_mag} amd \ref{mdbar_chi_mag}, 
as functions of the baryon density (in units of nuclear matter 
saturation density) for the values of magnetic field as
$eB=4 m_\pi^2$ and $eB=8 m_\pi^2$, with the values of the
isospin asymmetry parameter as $\eta$=0, 0.3 and 0.5,
with the anomalous magnetic moments (AMM) of the nucleons
taken into consideration. These results are compared with
the case when the AMM of the nucleons are neglected.
In the isospin symmetric nuclear matter, the masses (in MeV)
of the $D^+$ and $D^0$ are observed to be 1813.6 (1811.8) 
and 1787 (1785.5) 
at the nuclear matter saturation density for magnetic field 
(in units of 1/e) of $4 m_\pi^2$ with (without) AMM effects.
With the larger density pf 4$\rho_0$, these are modified to 
1557 (1542.6) and 1526.5 (1510.4).
These values may be compared to 1809.7 (1807.8) and 1782.9 (1781.4)
at $\rho_B=\rho_0$ and 1539.9 (1524.9) and 1508.5 (1492.7)
at $\rho_B=4\rho_0$,
in the frozen glueball approximation \cite{dmeson_mag}.
The modifications of the masses of the $D$ mesons are thus observed
to be rather small, when the medium modifications of the
dilaton field are taken into account, as compared to the
frozen glueball approximation.
%************************************************************
%Discussions about these masses as compared to those within the
%frozen chi approximation.
%************************************************************
For the $\bar D$ mesons, in isospin symmetric nuclear matter,
the masses of $D^-$ and $\bar {D^0}$ mesons are obtained as
1864 (1862.2) and 1837.8 (1836.4) at $\rho_0$ and 1740.2 (1722.7) and
1711.5 (1692.6) at 4$\rho_0$ with (without) AMM effects.  
The isospin asymmetry as well as AMM effects are observed 
to be larger at the higher magnetic field of $eB=8 m_\pi^ 2$
for $D^0$ meson, whereas $D^+$ is rather insensitive
to these effects, as can be seen from figure \ref{md_chi_mag}.
For   $D^-$ as well as $\bar {D^0}$ mesons, the asymmetry
as well as AMM effects are appreciable for the larger value
of the magnetic field ($eB=8 m_\pi ^2$), as can be seen
from figure \ref{mdbar_chi_mag}. 
%The behaviour of the $D$ and $\bar D$ mesons are observed to be
%similar to the case of frozen glueball approximation
%as has been discussed in detail in Ref. \cite{dmeson_mag}.
%%%%added discussions about D and D bar masses%%%%%%%%%%%%%%%%

%%%% added below%%%%%%%%%%%%%%%%%%%%%%%%%%%%%%%
The in-medium masses of the charmonium states,
$J/\psi$, $\psi(3686)$ and $\psi(3770)$ in the magnetized
asymmetric nuclear matter have been studied
within the chiral effective model \cite{charmonium_mag}.
In the present work, we study the mass modifications
of the P-wave charmonium states, $\chi_{c0}$ and $\chi_{c2}$,
and investigate the partial decay widths of the
charmonium states ($J/\psi, \psi(3686), \psi(3770),
\chi_{c0}, \chi_{c2}$) to $D\bar D$ in the magnetized
nuclear matter. 
%%%% added above%%%%%%%%%%%%%%%%%%%%%%%%%%%%%%%

In figures \ref{mchic0_mag} amd \ref{mchic2_mag}, the masses
of the 1P states are plotted as functions of the
baryon density (in units of nuclear matter saturation
density) for the values of magnetic field as
$eB=4 m_\pi^2$ and $eB=8 m_\pi^2$, with the values of the
isospin asymmetry parameter as $\eta$=0, 0.3 and 0.5,
with the anomalous magnetic moments (AMM) of the nucleons
taken into consideration. These results are compared with
the case when the AMM of the nucleons are neglected.
The mass of $\chi_{c0}$ (in MeV) is observed to be modified 
from its vacuum value of 3414.7 to
3385.57 (3309) at a density of $\rho_0 (4\rho_0)$
for $eB=4 m_\pi^2$ for symmetric nuclear matter,
with AMM effects, and 3384.8 (3303.7) when AMM
of nucleons are not taken into consideration.
The mass shifts are oberved to be smaller with isospin asymmetry 
effects, as well as with increase in the magnetic field.
The mass shifts are observed to be larger
when the AMM of the nucleons are not considered.
The mass of $\chi_{c2}$ is plotted in figure \ref{mchic2_mag}.
For symmetric nuclear matter, at a density of $\rho_0 (4\rho_0)$
for $eB=4 m_\pi^2$, the mass of $\chi_{c2}$ (in MeV) 
is observed to be modified 
from its vacuum value of 3556.17 to
3511 (3393) with AMM effects, and 3509.9 (3382.9) when AMM
of nucleons are not taken into consideration.
The effects of isospin asymmetry as well as magnetic fields
on the mass of $\chi_{c2}$ plotted in figure \ref{mchic2_mag}
are observed to be similar to that of $\chi_{c0}$
in the present work as
well as to the mass modifications of the charmonium
states $J/\psi$, $\psi(3686)$ and $\psi(3770)$
already studied in Ref. \cite{charmonium_mag}.
This is due to the fact that the mass modifications
of the charmonium states are determined by the
medium change of the scalar dilaton field
in the magnetized nuclear matter.
%%%%%%%%%comments regd comment 1 of Referee 1
The medium change of the mass of the charmonium state
is observed to be quite insensitive to the isospin asymmetry
in the medium. This can be understood in the following manner.
The mass shift of the charmonium state is obtained from the
shift of the dilaton field, $\chi$ and is proportional to the shift
in the fourth power of $\chi$ from the vacuum value
(as can be seen from equation (\ref{masspsi})), 
in the limit of massless quarks 
in the trace anomaly \cite{charmonium_mag} as considered 
in the present investigation.  
In the mean field approximation, the meson fields are replaced
by their expectation values and as has already been mentioned,
the value of $\chi$ is obtained through the solution of
coupled equations of the scalar fields, $\sigma$, $\zeta$,
$\delta$ and $\chi$ given by equations (\ref{sigma})--(\ref{chi}). 
In the presence of the isospin asymmetry
in the medium, i.e. for nonzero values of the isospin
asymmetry parameter, $\eta = (\rho_n-\rho_p)/(2\rho_B)$, 
the number density of the neutron is larger than the proton density
and the scalar densities of the proton and neutron become
different, with the value of $\delta$ being proportional
to their difference. The magnitudes of $\sigma$ and $\zeta$,
become larger in the asymmetric nuclear matter than in the 
symmetric nuclear matter case,
due to the contributions from $\delta^2$ term, as can be seen 
from the equations for $\sigma$ and $\zeta$ fields
given by (\ref{sigma}) and (\ref{zeta}).
However, due to the much smaller magntidue of $\delta$
as compared to $\sigma$ and $\zeta$ 
(e.g., $\delta \sim$ 2 MeV, $\sigma \sim$ 45 MeV,
$\zeta \sim$ 94 MeV at $\rho_B=2\rho_0$
for $eB=8 m_\pi^2$ and $\eta$=0.3), 
the modifications for $\sigma$ and
$\zeta$ are very small in the asymmetric nuclear matter 
as compared to symmetric nuclear matter. 
It might be noted here that in the absence of a magnetic field,
the scalar densities for the proton and neutron remain the same
and the value of $\delta$ is zero, whereas, in the presence of
a magnetic field, even for the case of isospin symmetry
in the medium ($\rho_p=\rho_n$), the scalar densities are different
for the proton and neutron (due to contribution of Landau energy
levels for proton in the presence of a magnetic field), 
which gives rise to a nonzero value (though small,
with a maximum value of about 0.5 MeV for $eB=8 m_\pi^2$) 
for the isovector scalar meson, $\delta$.
The dependence of $\chi$ on $\sigma$ and $\zeta$ 
(as can be seen from equation
(\ref{chi})) in the leading order, is proportional to  
$\big((\sigma^2-\delta^2)\zeta\big)$/($\sigma^2+\zeta^2+\delta^2$). 
The magnitude of the isovector scalar field $\delta$,
remains small ($\sim$ few MeV) as compared to the magnitudes 
of $\sigma$ or $\zeta$. This
is the reason why the modification of the 
dilton field $\chi$ due to isospin asymmetry remains
small, and subsequently the 
mass shifts of the charmonium states (which are proportional
to ($\chi^4 -\chi_0^4$)) remain insensitive to isospin
asymmetry of the medium.
%%%%%%%%%comments regd comment 1 of Referee 1 added above

We investigate the in-medium decay widths of the
charmonium states to $D\bar D$ in the magnetized
nuclear matter, for the values of the isospin asymmetry parameter
$\eta$ as 0, 0.3 and 0.5 and magnetic fields 
$eB=4 m_\pi^2$ and $eB=8 m_\pi^2$.
For $eB=8 m_\pi^2$ and in the case when the AMM effects
of the nucleons are neglected, the decay width 
of $J/\psi \rightarrow D^0 \bar {D^0}$ is observed 
to be possible above a density of
around 5$\rho_0$ in symmetric nuclear matter. 
There is observed to be an increase
in this decay width  with density,
reaching a value of around 27.9 MeV at 6$\rho_0$.
The decay $J/\psi \rightarrow D\bar D$
is not observed in any other  
case considered in the present work. 

The decay widths of $\psi(3686)$ to $D\bar D$
in magnetized isospin asymmetric nuclear matter 
are plotted as functions of the baryon density
in units of nuclear matter saturation density,
in figure \ref{psi3686_decay_mag} for different values
of the magnetic field and isospin asymmetry parameter.
These are plotted for the channels
(i) $D^+ D^-$, (ii) $D^0 \bar {D^0}$ as well as 
(iii) the sum of these two channels. The decay of $\psi (3686)$ 
is not possible in vacuum since the mass of this charmonium
state is smaller than the combined mass of $D$ and $\bar D$
in vacuum. However, at certain densities, these decays
become possible, when the center of mass momentum, $p_D$
becomes non-zero with the medium modifications of the masses of
the charmonium as well as the outgoing $D$ and $\bar D$ mesons
\cite{friman,amarvepja}. In symmetric nuclear matter
($\eta$=0), in the presence of magnetic field,
$eB=4 m_\pi^2$, shown in panel (a) of figure
\ref{psi3686_decay_mag}, the threshold  
densities above which the decays $\psi (3866)$ to $D^+D^-$
and $\psi (3866)$ to $D^0 \bar {D^0}$ become possible
are  4.5$\rho_0$ and 2.54$\rho_0$  respectively
when the AMM of the nucleons are taken into account,
and 4.1$\rho_0$ and 2.37$\rho_0$
when the AMM of nucleons are not considered.
The higher value for the threshold
density for the decay width of $\psi (3866)$ to $D^+D^-$,
as compared to the channel  $\psi (3866)$ to $D^0 \bar {D^0}$
is due to the reason that 
the masses of the charged $D$ as well as charged $\bar D$ mesons
have positive shifts due to the contributions from the 
lowest Landau levels \cite{dmeson_mag}, which makes 
the mass of $D^+ D^-$ to be larger than $D^0 \bar {D^0}$.
For the magnetic field $eB=8 m_\pi^2$
in symmetric nuclear matter, as shown in panel (b),
the decay of
$\psi (3866)$ to $D^0 \bar {D^0}$ becomes possible
at a density of 2.4$\rho_0$ (2.2$\rho_0$)
for the cases of with (without) AMM effects.
The decay of
$\psi (3866)$ to $D^+ {D^-}$ is not observed
even upto a density of 6$\rho_0$.

In the isospin asymmetric nuclear medium
with $\eta$=0.3, the densities
above which the decays of $\psi (3866)$ to $D^+D^-$
and $\psi (3866)$ to $D^0 \bar {D^0}$ become possible 
are observed to be 3.3$\rho_0$ and 3.7$\rho_0$ respectively
for $eB=4 m_\pi^2$ as can be seen from panel (c) 
of figure \ref{psi3686_decay_mag} when the AMM
of the nucleons are taken into account.  
These densities are modified to 3.12$\rho_0$ and 3.5$\rho_0$,
when the AMM effects are not taken into consideration.
The contrasting behaviour of
the threshold density for $\psi (3686) \rightarrow D^+ D^-$
to be smaller than the channel 
$\psi (3686) \rightarrow D^0 \bar {D^0}$
for the isospin asymmetric case with $\eta$=0.3,
as compared to the isospin symmetric case 
for $eB=4 m_\pi^2$, is due to the reason that
the center of mass momentum of the outgoing
$D$ and $\bar D$ mesons, $p_D$, has larger
contribution from the last term of equation
(\ref{pd}) for the case of outgoing mesons
as $D^+ D^-$ as compared to $D^0 \bar {D^0}$, 
since the difference in the masses
of $D^+$ and $D^-$ is much bigger than 
the difference in $D^0$ and $\bar {D^0}$
masses. 
When the magnetic field is increased to 
$eB=8 m_\pi^2$, for asymmetric nuclear matter with $\eta$=0.3, 
as can be seen from panel (d) of figure \ref{psi3686_decay_mag}, 
the values for the threhold densities 
for decay to $D^+ D^-$ and $D^0 \bar {D^0}$ 
are 4.6$\rho_0$ and 3.6$\rho_0$ respectively for the case
without AMM effects. The higher threhold density 
for the charged $D\bar D$
decay as compared to neutral $D\bar D$ decay is similar
to the case of the isospin symmetric case.
This is because the masses of the charged mesons have larger 
positive mass shifts with increase in the magnetic field
to $eB=8 m_\pi^2$.
When the AMM effects are taken into account,
the threshold dnesity for $\psi (3866)$ to $D^0 \bar {D^0}$ 
is about 4$\rho_0$ and the decay of $\psi (3866)$ to 
$D^+D^-$ does not become possible even upto a density of 6$\rho_0$.

With further increase in the isospin asmmetry, with $\eta$=0.5,
only the decay channel of $\psi(3686)$ to $D^+ D^-$  
is possible, above a density of 2.9$\rho_0$ and 5$\rho_0$
for $eB=4 m_\pi^2$ and $eB=8 m_\pi^2$ respectively,
when AMM of the nucleons are considered. In the case
when the AMM effects are not taken into account,
there is no dependence on the magnetic field,
for the decay to $D^0 \bar {D^0}$,
as the system consists of only neutrons for $\eta$=0.5
and the only effect of magnetic field can be through the
AMM of neutron. However, in the presence of a magnetic field,
the masses of the charged $D^\pm$ mesons are modified
due to the contribution from the lowest Landau level.
For the case of without AMM effects
and for $\eta$=0.5, the decay of $\psi (3686)$ to 
$D^+D^-$ is possible above a density of about 2.7 (4)$\rho_0$
for $eB=4 (8) m_\pi^2$,
and the decay to the neutral $D\bar D$ pair is not observed
even upto a density of 6$\rho_0$. 

In figure \ref{psi3770_decay_mag}, 
the partial decay widths of $\psi (3770)$ to $D\bar D$
are shown for different values of the isospin asmmetry
parameter and magnetic fields, accounting for the effects
of AMMs of nucleons. These results are compared with
the case when the AMMs of nucleons are not taken into
consideration.
For symmetric nuclear matter ($\eta$=0), there is observed to be
an initial decrease of the decay width upto a density
of around 1.5$\rho_0$ followed by an increase 
at high densities both for the cases of 
$eB=4 m_\pi^2$ and $eB=8 m_\pi^2$, as can be seen
from panels (a) and (b) of figure \ref{psi3770_decay_mag}. 
For $eB=4 m_\pi^2$, the decay is solely through the channel
$\psi(3770) \rightarrow D^0 \bar {D^0}$ upto 
a density of around 4.8$\rho_0$ (5.3$\rho_0$),
above which the decay to the charged $D\bar D$ mesons 
also becomes possible. For the higher magnetic field of
$eB=8 m_\pi^2$, as can be seen from panel (b)
of figure \ref{psi3770_decay_mag}, 
the decay to the charged $D\bar D$ mesons
does not become possible even upto a density
of 6$\rho_0$.
For isospin asymmetric nuclear matter with $\eta$=0.3,
there is observed to be an intial drop in the partial
decay width of $\psi(3770)\rightarrow D\bar D$
(solely due to contribution from the channel 
$\psi (3770) \rightarrow D^0 \bar {D^0}$) 
with density, which becomes very small 
($\sim$ 0.28 (0.11) MeV)
at around 1.05$\rho_0$ (0.9$\rho_0$) with (without) 
the AMM effects of the nucleons. 
The value of the decay width for
$\psi (3770) \rightarrow D^0 \bar {D^0}$ 
remains similar upto a density of around 3.75$\rho_0$
(3.6$\rho_0$), with (without) AMM effects,
above which there is observed to be an increase
in this decay width with density.
The decay channel
$\psi (3770) \rightarrow D^+ {D^-}$ is observed to 
become possible above a density of around 
3.1$\rho_0$ (2.9$\rho_0$) when the AMM effects 
are considered (neglected).
For the magnetic field $eB=8m_\pi^2$, the behaviour
is similar to case of $eB=4m_\pi^2$, for the case
when the AMMs of the nucleons are not taken into account.
On the other hand, for the case when AMM effects 
are considered, there is no decay to charged
$D\bar D$ pair observed even upto a density of 6$\rho_0$.
For the isospin asymmetry parameter, $\eta$=0.5,
the decay is only through the channel  
$\psi (3770) \rightarrow D^+ D^-$, for both
$eB=4m_\pi^2$ and $eB=8m_\pi^2$.

In figure \ref{chic0_decay_mag}, 
the effects of isospin asymetry, magnetic field and density
on the partial decay widths of $\chi_{c0}$ to $D\bar D$
are shown. There is observed to be a threshold 
density above which the decay is observed to be
possible for both the charged and neutral $D\bar D$ channels.
There is observed to be an initial increase with density,
followed by a drop when the density is further increased.
The AMM effects are observed to be large
for higher value of magnetic field.

In figure \ref{chic2_decay_mag}, 
the partial decay widths of $\chi_{c2} \rightarrow D\bar D$
is plotted as functions of the baryon density for
different values of isospin asymmetry parameter
and magnetic fields, both with and without AMM effects
of the nucleons. There is observed to be a threshold 
density for both the charged and neutral $D\bar D$ pair
channels, and the decay width is observed to increase 
monotonically with density. This behaviour can be understood
looking at the polynomial part (in $p_D$) 
of the partial decay width which increases with $p_D$,
which in turn is an increasing function of the baryon density.
This behaviour of monotonic increase in the decay width of
$\chi_{c2}$ was also observed in Ref. \cite{friman},
as a function of the mass drop in $D$ and $\bar D$
mesons in the nuclear medium.

%%%%added discussions regd comment 2 of Referee 2
The partial decay widths of the charmonium to $D\bar D$ 
depend on the center of mass momentum of the $D(\bar D)$
meson, $p_D$ (given by equation (\ref{pd}),
through a polynomial part multiplied by an exponential
part, as can be seen from the equations 
(\ref{jpsidw})--(\ref{chic2dw}). 
The stark difference of the  $\chi_{c0}$ and $\chi_{c2}$ 
decay widths is due to the difference in the forms 
of the polynomial part,
which has a monotonic increase with density in the case
of $\chi_{c2}$ above a threshold density when the decay to
$D\bar D$ becomes possible. On the other hand, the
decay width for $\chi_{c0}$ (above the threshold density
when the decay becomes possible) shows an initial 
increase followed by a drop with further increase
in the density. Similar behaviour for the decay wdiths
of $\chi_{c0}$ and $\chi_{c2}$ to $D\bar D$ were also
observed in Ref. \cite{friman}. 
The medium modifications of these decay widths were 
calculated from the mass modifications of the $D$ and
$\bar D$ mesons, which were assumed to be same in the
symmetric nuclear matter in Ref. \cite{friman}. 
The $\psi(3686)$ decay width shows 
a monotonic increase, whereas $\psi(3770) \rightarrow D\bar D$
decay width has an initial drop followed by a rise 
with further increase in the density. As has already been
mentioned, the dependence of the charmonium decay width
to $D\bar D$ on density is determined by the form of the 
polynomial and the exponential parts in terms of the
center of mass momentum, $p_D$.
%%%%added discussions regd comment 2 of Referee 2

\section{Summary}
We have studied the effects of magnetic field on the
partial decay widths of the charmonium states,
$J/\psi (3097)$, $\psi(3686)$, $\psi (3770)$, $\chi_{c0}$
and $\chi_{c2}$ to $D\bar D$ in isospin asymmetric nuclear
matter. These are obtained using the mass modifications
of the $D$ and $\bar D$ mesons as well as the charmonium
states, calculated within a chiral effective model.
The mass modifications of the open charm mesons,
$D$ ($D^0$,$D^+$) and $\bar D$ ($\bar {D^0}$, $D^-$)
arise due to their interactions with the nucleons
and the scalar mesons, where as the in-medium masses
of the charmonium states are calculated from the medium
changes of the scalar dilaton field, which simulates
the gluon condesates of QCD within the chiral effective model.
In the presence of the magnetic field, the proton has
contributions from the Landau energy levels.
There is observed to be unequal masses for the
$D^0$ and $D^+$  within the $D$ meson doublet,
as well as for $\bar {D^0}$ and $D^-$ within the
$\bar D$ doublet, even in isospin symmetric nuclear matter,
due to difference in the interactions with the proton and neutron,
as well as due to contributions from the Landau energy levels
for the charged $D$ and $\bar D$ mesons,
in the presence of magnetic field. 
In symmetric nuclear matter, the threshold densities 
for which the decay widths
of the charmonium state to $D\bar D$ turn out to be smaller
for the $D^0 \bar {D^0}$ channel as compared
to for $D^+ D^-$ channel, as the masses of the
$D^+$ as well as $D^-$ have positive shifts 
in the presence of magnetic field.
The effects of isospin asymmetry, magnetic fields
are observed to be quite appreciable on the
partial decay widths for the charged and neutral
$D\bar D$ pair channels, which will affect the
production of these mesons in high energy 
asymmetric heavy ion collision experiments.  

\acknowledgements
One of the authors (AM) is grateful to ITP, University of Frankfurt,
for warm hospitality and 
acknowledges financial support from Alexander von Humboldt Stiftung 
when this work was initiated. 
Amal Jahan CS acknowledges the support
towards this work from Department of Science and Technology, Govt of India,
via INSPIRE fellowship scheme (INSPIRE Code IF170745). 

%\begin{references}

%\end{references}
\end{document}